\documentclass[11pt]{article}
\usepackage[a4paper,margin=25mm]{geometry}
\usepackage[T1]{fontenc}
\usepackage[utf8]{inputenc}
\usepackage{lmodern}
\usepackage{microtype}
\usepackage{amsmath,amssymb}
\usepackage{graphicx}
\usepackage{booktabs}
\usepackage{array}
\usepackage{enumitem}
\usepackage{hyperref}
\usepackage{xcolor}
\usepackage{caption}
\usepackage{placeins}
\usepackage{algorithm}
\usepackage{algpseudocode}
\usepackage{url}
\hypersetup{colorlinks=true,linkcolor=black,citecolor=black,urlcolor=blue}
\setlist[itemize]{leftmargin=*,nosep}
\setlist[enumerate]{leftmargin=*,nosep}
\title{\textbf{Governed Human--AI Prioritization Under Uncertainty:}\\
Adaptive Estimation and Dependency-Constrained Portfolio Selection}
\author{Azzeddine Ihsine \quad Sara Ihsine\\[2pt]
\small Inovionix\\[2pt]
\small \texttt{azzeddine.ihsine@inovionix.com, sara.ihsine@inovionix.com}}
\date{}

\begin{document}
\maketitle

\begin{abstract}
AI-native software engineering increasingly combines human judgment, historical analogy, parametric estimation, and AI-generated forecasts inside the same prioritization decision. We study five quantitative operators used in the D-POAF decision practice and stress-test their behavior under uncertainty, estimator dependence, drift, shared context error, dependencies, and limited capacity. All robustness extensions are repeated across 30 deterministic seeds. Ranking experiments confirm that moderate scoring changes preserve broad order while materially changing funded-set membership. Under independent estimator errors, inverse-MSE aggregation reaches mean MAE $0.615$, a $42.2\%$ reduction relative to the best individual estimator. That advantage contracts to $4.4\%$ at error correlation $\rho=0.6$ and becomes negative at $\rho=0.8$, while calibration-set regression remains slightly better than the best individual and closely approaches a test-set oracle. Under estimator drift, four-Wave adaptive weighting reduces average RMSE by $4.64\%$ (95\% CI $[4.50\%,4.78\%]$). Model--collective divergence achieves ROC-AUC $0.908$ with independent channels and remains above $0.878$ across shared-noise conditions up to $0.8$. We separate budget-constrained portfolio selection from ODP's native dependency-aware sequencing role. When ODP distance is repurposed as a portfolio-selection heuristic, value-to-effort is stronger in the tested generator. For a fixed selected set, however, ascending $\mathrm{ERS}/\mathrm{BVS}$ is provably optimal for an ERS-weighted completion objective without dependencies and achieves mean efficiency $0.982$ relative to the exact precedence-constrained optimum at dependency intensity $\lambda=1.0$. In scaling to 100 Feature Blocks, ODP has the highest mean early-value AUC among the transparent sequencing baselines, although its margin over PVS ordering is small. These results characterize not only when evidence-weighted human--AI prioritization works, but also the boundary between selection and sequencing objectives.
\end{abstract}


\section{Introduction}
Software engineering decisions routinely trade expected value against effort, risk, timing, dependencies, and limited capacity. Value-Based Software Engineering established that software work is not value-neutral and that economic reasoning belongs inside engineering decisions~\cite{boehm2003}. Requirements prioritization reached the same problem from another direction: cost--value methods compare stakeholder value and implementation cost~\cite{karlsson1997}, while the Next Release Problem formalizes release selection under resource limits and demonstrates that realistic selection is computationally non-trivial~\cite{bagnall2001}. Reviews of prioritization methods report a broad landscape of ranking, voting, optimization, search-based, fuzzy, and learning approaches, together with persistent challenges around dependencies, uncertainty, stakeholder coordination, and empirical evaluation~\cite{achimugu2014,pitangueira2015,bukhsh2020}.

AI-native engineering intensifies this decision problem because multiple estimators can influence the same commitment. A domain expert may estimate from experience, an analogy method from similar work, a parametric model from project attributes, and an AI system from learned patterns. Software-effort research has long shown strong context dependence across estimation families~\cite{jorgensen2004,jorgensen2007}. Ensemble reviews show that combining estimators frequently improves accuracy, while dynamic selection remains comparatively underexplored and no single ensemble rule dominates every context~\cite{cabral2023}. Human--AI decision research adds a second requirement: predictive accuracy alone does not determine the quality of a human--AI team, and poorly calibrated decision support can produce overreliance~\cite{bansal2021,bucinca2021}.

This paper addresses four linked scientific problems. First, how stable are explicit prioritization scores when their strategic parameters change? Second, can observed predictive reliability determine estimator influence more effectively than fixed role-based weighting? Third, when dependencies and budgets turn release selection into a combinatorial portfolio problem, how close do transparent heuristics come to the exact optimum? Fourth, once a set of Feature Blocks has been selected, does the ODP distance $\mathrm{ERS}/\mathrm{BVS}$ produce efficient dependency-valid execution sequences?

We instantiate these questions using the decision operators documented in the D-POAF reference practice~\cite{dpoafcanonical,dpoafterminology,dpoafguide}. D-POAF supplies the governance context, but the quantitative contribution is independent of the framework label: the experiments test parameterized ranking, reliability-weighted estimation, calibration signals, dependency-constrained portfolio selection, and dependency-aware sequencing as explicit computational mechanisms. Selection and sequencing are evaluated separately because they optimize different objects.

Our contributions are:
\begin{enumerate}
  \item a formal separation of expected value, feasibility, calibration, and post-delivery value, preventing one score from silently standing in for all four;
  \item a 30-seed regeneration of BVS and PVS sensitivity showing that global rank stability can coexist with material funded-set changes;
  \item a correlation stress test of effort aggregation that compares inverse-MSE, calibration-set regression, equal weighting, individual estimators, and a test-set oracle ceiling;
  \item a multi-seed drift and shared-noise evaluation that quantifies both the benefit and the failure boundaries of adaptive reliability weighting and model--collective divergence;
  \item an exact portfolio benchmark plus a binary mixed-integer linear programming (MILP) reference scaled to 100 candidates, establishing how transparent heuristics behave as dependency-constrained selection grows;
  \item a native ODP sequencing benchmark that proves the $\mathrm{ERS}/\mathrm{BVS}$ ratio rule for the dependency-free weighted-completion objective, then measures its gap to exact precedence-constrained sequence optima and its scaling behavior to 100 Feature Blocks.
\end{enumerate}

\section{Related Work}
\subsection{Value and requirements prioritization}
Value-Based Software Engineering integrates stakeholder value, risk, and investment reasoning into software practice~\cite{boehm2003}. Karlsson and Ryan's cost--value approach treats prioritization as an explicit value--cost trade-off rather than a single importance rank~\cite{karlsson1997}. The Next Release Problem formalizes selection under budgets and customer demand, and its complexity result shows why realistic release planning cannot generally be reduced to sorting~\cite{bagnall2001}. Search-Based Software Engineering has extended this line to single- and multi-objective optimization, while reviews continue to identify risk, dependencies, uncertainty, scalability, and empirical validation as active concerns~\cite{pitangueira2015,bukhsh2020}.

The contribution studied here is a governed decomposition of the decision: value, feasibility, calibration, dependencies, budget, and measured outcomes remain separately traceable rather than being collapsed into an opaque ranking procedure.

Selection and sequencing are also distinct optimization problems. In single-machine scheduling without precedence constraints, Smith's ratio rule orders jobs by processing time divided by weight to minimize total weighted completion time~\cite{smith1956}. This classical result provides a precise experimental lens for the ODP distance $\mathrm{ERS}/\mathrm{BVS}$ when a release set is fixed, while prerequisite constraints remove the unconstrained optimality guarantee.

\subsection{Effort estimation and uncertainty}
Expert judgment remains central to software estimation. Evidence-based guidance recommends structured judgment, historical data, independent estimates, explicit uncertainty assessment, and feedback on prior accuracy~\cite{jorgensen2004,jorgensen2005}. A systematic review of software cost-estimation research found strong context dependence across approaches~\cite{jorgensen2007}. Ensemble effort estimation seeks to reduce individual-model weaknesses; an updated review reports frequent ensemble gains and comparatively limited work on dynamic ensemble selection~\cite{cabral2023}. This motivates reliability-weighted aggregation in which estimator influence is updated from measured predictive performance.

\subsection{Human--AI calibration}
Human--AI collaboration adds a calibration problem beyond raw prediction accuracy. Bansal et al. show that the most accurate standalone AI is not necessarily the best teammate~\cite{bansal2021}, while Bucinca et al. demonstrate that AI-assisted decisions can induce overreliance and that interventions can alter this behavior~\cite{bucinca2021}. In the model studied here, collective judgment does not replace predictive scoring; it provides an independent calibration signal capable of triggering investigation when model and collective assessments diverge.

\section{Decision Model}
\subsection{Business Value Score}
For a candidate block, BVS aggregates three $1$--$10$ judgments:
\begin{equation}
\mathrm{BVS}=w_1 I_m+w_2 U_m+w_3 O_s,
\end{equation}
with $w_i\geq0$ and $\sum_i w_i=1$. $I_m$ denotes business impact, $U_m$ urgency or criticality, and $O_s$ strategic opportunity. The weights are explicit strategy parameters whose changes can be versioned and replayed~\cite{dpoafguide}.

\subsection{Effort and Risk Score}
The feasibility score is
\begin{equation}
\mathrm{ERS}=1+\beta E+\gamma R,
\end{equation}
where $E$ is estimated effort, $R$ technical risk, and $\beta,\gamma$ govern sensitivity to each factor. The reference practice also permits multiple estimation sources and explicit uncertainty adjustment~\cite{dpoafguide}.

\subsection{Prioritization Value Score}
Prioritization combines value and feasibility through
\begin{equation}
\mathrm{PVS}_{\alpha}=\frac{\mathrm{BVS}^{\alpha}}{\mathrm{ERS}}.
\end{equation}
The exponent $\alpha$ controls how aggressively high business value dominates feasibility. Rather than treating $\alpha$ as a hidden tuning constant, we measure how its changes affect ranking and top-portfolio membership.

\subsection{Collective calibration}
The D-POAF reference practice defines a Collective Calibration Score using computed priority and allocated voting points~\cite{dpoafguide}. Our experiment isolates the diagnostic principle: if a model estimate and an independently generated collective signal diverge strongly, that divergence can be used to identify candidate miscalibration before commitment. Calibration and realized value remain separate constructs.

\subsection{ODP, sequencing, and the optimization boundary}
The reference practice assigns each Feature Block the distance
\begin{equation}
d_i=\frac{\mathrm{ERS}_i}{\mathrm{BVS}_i},
\end{equation}
and uses dependency-aware path reasoning to sequence work~\cite{dpoafguide}. This role must be distinguished from selecting which blocks enter a budget-constrained release. The latter is a subset-selection problem. ODP is primarily a sequencing model over blocks whose value, effort/risk, and prerequisites are already explicit.

To test that native sequencing role, let $\pi$ be a dependency-valid order over a fixed selected set. Define the ERS completion burden of block $i$ as
\begin{equation}
C_i^{\mathrm{ERS}}(\pi)=\sum_{j:\,\pi(j)\leq \pi(i)} \mathrm{ERS}_j,
\end{equation}
and define the sequencing objective
\begin{equation}
J(\pi)=\sum_i \mathrm{BVS}_i C_i^{\mathrm{ERS}}(\pi).
\end{equation}
Lower $J$ means that high-value blocks become available earlier under cumulative effort/risk burden. For two unconstrained adjacent blocks $i$ and $j$, the difference between ordering $i$ before $j$ and $j$ before $i$ is
\begin{equation}
J_{ij}-J_{ji}=\mathrm{BVS}_j\mathrm{ERS}_i-\mathrm{BVS}_i\mathrm{ERS}_j.
\end{equation}
Thus $i$ should precede $j$ exactly when $\mathrm{ERS}_i/\mathrm{BVS}_i\leq \mathrm{ERS}_j/\mathrm{BVS}_j$. This is Smith's ratio rule~\cite{smith1956}. Hence ascending ODP distance is optimal for this declared objective when no prerequisite constraints exist. With prerequisites, selecting the lowest-distance currently eligible block remains dependency-valid but is not guaranteed globally optimal, so we benchmark it against an exact precedence-constrained dynamic program.

Algorithm~\ref{alg:odp} states the dependency-aware ODP policy evaluated in the sequencing experiments. The input set is first closed under prerequisites. In the benchmark, every generated prerequisite already belongs to the fixed selected set, so this closure is a no-op. At each step, ODP chooses the currently eligible Feature Block with the smallest $\mathrm{ERS}/\mathrm{BVS}$ distance. A deterministic identifier tie-break makes the policy reproducible.

\begin{algorithm}[t]
\caption{ODP dependency-aware sequencing}
\label{alg:odp}
\begin{algorithmic}[1]
\Require Selected Feature Blocks $S$, prerequisite sets $P(i)$, $\mathrm{BVS}_i>0$, $\mathrm{ERS}_i>0$
\State $S \gets \mathrm{TransitiveDependencyClosure}(S)$
\ForAll{$i \in S$}
    \State $d_i \gets \mathrm{ERS}_i/\mathrm{BVS}_i$
\EndFor
\State $D \gets \emptyset$; $\pi \gets [\,]$
\While{$|D| < |S|$}
    \State $A \gets \{i \in S\setminus D : P(i) \subseteq D\}$
    \If{$A=\emptyset$}
        \State \Return \textsc{Infeasible} \Comment{cycle or missing prerequisite}
    \EndIf
    \State $i^* \gets \arg\min_{i\in A}(d_i,\,i)$ \Comment{identifier breaks ties}
    \State append $i^*$ to $\pi$; $D \gets D \cup \{i^*\}$
\EndWhile
\State \Return $\pi$
\end{algorithmic}
\end{algorithm}

Separately, we retain the earlier portfolio experiment in which ODP distance is intentionally repurposed as a transparent heuristic for budget-constrained release selection. That experiment tests an optimization boundary rather than ODP's native role. General release selection remains a different combinatorial problem~\cite{bagnall2001,pitangueira2015}.

\section{Research Questions and Experimental Method}
We evaluate five research questions:
\begin{itemize}
  \item \textbf{RQ1:} How sensitive are BVS, ERS, and PVS rankings to governed parameters across independent simulation replications?
  \item \textbf{RQ2:} How does estimator-error correlation change the performance of individual, inverse-MSE, regression-fitted, equal, and oracle aggregation?
  \item \textbf{RQ3:} Do adaptive reliability weighting and model--collective divergence remain useful under nonstationarity and shared error?
  \item \textbf{RQ4:} When ODP distance is repurposed as a budget-constrained release-selection heuristic, how close do transparent selection heuristics come to the exact portfolio optimum as candidate count grows?
  \item \textbf{RQ5:} For ODP's native role, how efficiently does $\mathrm{ERS}/\mathrm{BVS}$ sequence a fixed set of Feature Blocks under prerequisite constraints, and how close is it to the exact precedence-constrained sequence optimum?
\end{itemize}

The robustness extension uses 30 deterministic seeds derived from master seed 20260827 and NumPy's \texttt{Generator(PCG64)} random-number generator. The source archive records every replication seed, software version, generated CSV, and figure script. Confidence intervals reported for the new studies treat the seed as the replication unit rather than individual tasks or portfolio instances.

\subsection{Ranking sensitivity}
For each seed, 5,000 candidate blocks receive three independent $1$--$10$ value components and baseline BVS weights $(0.4,0.3,0.3)$. BVS weight perturbations are sampled from Dirichlet distributions centered on those weights with concentration $\kappa\in\{12,24,60\}$, where larger $\kappa$ means narrower perturbation. Fifty weight draws are evaluated per $\kappa$ and seed. PVS sensitivity uses effort sampled from a bounded lognormal distribution, risk sampled uniformly on $[1,10]$, $\mathrm{ERS}=1+0.15E+0.15R$, and $\alpha\in\{1,1.5,2\}$.

\subsection{Correlated estimator errors and aggregation baselines}
The original experiment used independent estimator errors. To isolate that assumption without changing each estimator's marginal quality, the revised generator fixes the four marginal biases and RMSE values to those observed in the original calibration study and introduces an equicorrelated error component. For estimator $i$ and task $j$,
\begin{equation}
e_{ij}=b_i+\sigma_i\left(\sqrt{\rho}\,z_{0j}+\sqrt{1-\rho}\,z_{ij}\right),
\end{equation}
where $z_{0j}$ and $z_{ij}$ are independent standard normal variables, $b_i$ is the fixed estimator bias, and $\sigma_i=\sqrt{\mathrm{RMSE}_i^2-b_i^2}$. Sweeping $\rho$ from $0$ to $0.8$ holds each marginal error scale approximately fixed while increasing shared error. Every seed uses 2,500 calibration tasks and 10,000 disjoint test tasks.

Inverse-MSE weighting sets $w_i\propto1/\widehat{\mathrm{MSE}}_i$ on the calibration set. A second deployable baseline fits nonnegative simplex weights by constrained regression,
\begin{equation}
\min_{w}\frac{1}{n}\sum_{j=1}^{n}\left(\sum_i w_i e_{ij}\right)^2,
\qquad w_i\geq0,\quad \sum_i w_i=1.
\end{equation}
An oracle applies the same optimization directly to the test errors and is used only as an unattainable ceiling. Equal weighting and all four individual estimators remain baselines.

\subsection{Drift and shared-noise calibration}
The drift experiment uses 20 Waves with 250 tasks each. Expert, analogy, and parametric error scales remain fixed while AI error standard deviation improves linearly from $2.0$ to $0.9$. Adaptive inverse-MSE weights are recomputed from the previous \textbf{four Waves}, or 1,000 tasks once the window is full. The first Wave uses equal weights. The complete experiment is repeated over all 30 seeds.

For CCS, each seed generates 6,000 proposals. Model error contains unit Gaussian noise and an additional signed stale-context shift of magnitude $1.5$ for 15\% of proposals. Collective error has scale $0.6$ and shares a tunable fraction $\rho_s$ of the model's base noise:
\begin{equation}
e_c=0.6\left(\rho_s z_m+\sqrt{1-\rho_s^2}\,z_c\right),\qquad \rho_s\in[0,0.8].
\end{equation}
The diagnostic is absolute model--collective divergence. Positives are the top 20\% of absolute model errors. We report ROC-AUC and the lift in precision obtained by flagging the top 20\% by divergence. Recall is omitted because, under equal flagged and positive fractions, it is numerically identical to precision by construction.

\subsection{Portfolio scaling with an exact binary MILP reference}
The published 800-instance, 12-candidate benchmark is retained. We add a fresh scaling family at $n\in\{12,25,50,100\}$ with five instances per seed and size, giving 150 instances per size. BVS is sampled from a normal distribution truncated to $[1,10]$, effort from a bounded lognormal distribution, risk uniformly from $[1,10]$, and each candidate receives zero, one, or two prerequisites from earlier candidates with probabilities $0.45$, $0.40$, and $0.15$. The effort budget equals 30\% of total candidate effort.

The exact selection reference is a binary mixed-integer linear program (MILP), equivalently a 0--1 integer linear program in this formulation. Let $x_i\in\{0,1\}$ indicate whether Feature Block $i$ is selected, let $v_i=\mathrm{BVS}_i$, let $c_i$ be its effort cost, let $B$ be the release effort budget, and let $\mathcal{D}$ contain ordered pairs $(i,j)$ meaning that block $i$ requires prerequisite $j$. The benchmark solves
\begin{align}
\max_{x}\quad & \sum_{i=1}^{n} v_i x_i, \\
\text{s.t.}\quad & \sum_{i=1}^{n} c_i x_i \leq B, \\
& x_i \leq x_j, && \forall (i,j)\in\mathcal{D}, \\
& x_i \in \{0,1\}, && i=1,\ldots,n.
\end{align}
The first constraint enforces the release budget. The dependency constraint makes a candidate selectable only when every declared prerequisite is also selected. The exact objective intentionally matches the declared portfolio-selection question, maximize total BVS under an effort budget and prerequisites. Risk is generated because ODP-distance and PVS baselines use ERS, but it is not an additional constraint in this exact selection objective. The model is solved through SciPy's MILP interface backed by HiGHS~\cite{scipy2020}.

Algorithm~\ref{alg:portfolio} summarizes the exact-reference procedure and its small-instance validation. For one $n=12$ instance per seed, the MILP optimum is independently checked by exhaustive enumeration over all $2^n$ subsets. All 30 checks agree to numerical tolerance, with maximum absolute difference $1.5\times10^{-14}$.

\begin{algorithm}[t]
\caption{Exact dependency-constrained portfolio benchmark}
\label{alg:portfolio}
\begin{algorithmic}[1]
\Require Values $v_i$, effort costs $c_i$, budget $B$, dependencies $\mathcal{D}$
\State create binary decision variables $x_i\in\{0,1\}$ for $i=1,\ldots,n$
\State add budget constraint $\sum_i c_i x_i \leq B$
\ForAll{$(i,j)\in\mathcal{D}$}
    \State add prerequisite constraint $x_i \leq x_j$
\EndFor
\State $x^* \gets \mathrm{MILPSolve}(\max \sum_i v_i x_i)$
\State $V^* \gets \sum_i v_i x_i^*$
\If{$n=12$ and instance is designated for validation}
    \State $V_{\mathrm{enum}} \gets \mathrm{ExhaustiveFeasibleMaximum}(v,c,B,\mathcal{D})$
    \State assert $|V^*-V_{\mathrm{enum}}|\leq\varepsilon$
\EndIf
\State \Return $V^*$
\end{algorithmic}
\end{algorithm}

\subsection{Native ODP sequencing benchmark}
The sequencing benchmark fixes the selected Feature Blocks and asks only for a dependency-valid execution order. This deliberately removes the budget-selection decision from the experiment. BVS, effort, and risk are generated from the same families as the scaling study, then $\mathrm{ERS}=1+0.10E+0.10R$. Dependency DAGs are generated by allowing each block to draw up to three prerequisites from earlier blocks. A parameter $\lambda$ controls the expected direct prerequisite count before that cap.

For the exact study, $n=12$, $\lambda\in\{0,0.5,1.0,1.5\}$, and eight instances are generated per seed and dependency level. We compare five dependency-valid policies: ODP, which selects the currently eligible block with minimum $\mathrm{ERS}/\mathrm{BVS}$, PVS $\alpha=2$, BVS-first, ERS-first, and random topological ordering. An exact dynamic program enumerates feasible completion subsets and minimizes $J(\pi)=\sum_i \mathrm{BVS}_i C_i^{\mathrm{ERS}}(\pi)$. We report sequence efficiency $J^*/J$, where one is exact optimality.

A second study fixes $\lambda=1.0$ and scales to $n\in\{12,25,50,100\}$ with five instances per seed and size. Because the selected set is fixed, every policy eventually delivers the same total BVS. Sequence quality is therefore measured by when that value becomes available. We report normalized early-value area
\begin{equation}
\mathrm{AUC}_{V}=\frac{VT-J(\pi)}{VT},\qquad V=\sum_i\mathrm{BVS}_i,\quad T=\sum_i\mathrm{ERS}_i,
\end{equation}
which is equivalent to the area under cumulative delivered BVS against cumulative ERS, together with the ERS fraction required to deliver 50\% and 80\% of total BVS. All generated sequences are checked for prerequisite validity.

\begin{table}[t]
\caption{Revised experimental design and supported inference.}
\centering\small
\begin{tabular}{p{0.18\columnwidth}p{0.74\columnwidth}}
\toprule
Study & Test and inference \\
\midrule
Sensitivity & Thirty-seed rank and top-10\% stability under BVS and PVS parameter changes. \\
Effort & Correlation sweep $\rho=0$ to $0.8$, four aggregation rules, four individual estimators, and an oracle ceiling. \\
Adaptation & Equal versus four-Wave adaptive reliability weighting under improving AI error. \\
CCS & Diagnostic AUC and lift as model and collective errors share increasing context noise. \\
Portfolio selection & Published 800-instance exact benchmark plus MILP-referenced scaling to $n=100$, with ODP distance treated explicitly as a selection heuristic rather than its native role. \\
ODP sequencing & Exact $n=12$ weighted-completion benchmark across four dependency intensities plus dependency-valid scaling to $n=100$. \\
\bottomrule
\end{tabular}
\end{table}

\subsection{Evaluation metrics}
Rank stability uses Spearman correlation and top-10\% overlap. Effort prediction uses MAE and RMSE, with relative MAE gain reported against the best individual estimator. CCS diagnostic quality uses ROC-AUC and top-20\% precision lift over the 20\% base rate. Portfolio-selection quality uses objective ratio relative to the exact optimum, normalized regret, and exact-optimum rate. Native sequencing uses efficiency $J^*/J$ for the exact study, normalized early-value AUC, and ERS fractions to 50\% and 80\% delivered BVS. For new robustness experiments, 95\% confidence intervals are computed over the 30 seed-level replications with Student-$t$ intervals. The original 800-instance portfolio benchmark retains its nonparametric bootstrap intervals.
\section{Results}
\subsection{RQ1: governed parameters materially shape the top portfolio}
The 30-seed regeneration reproduces the ranking-sensitivity pattern. Under narrow BVS perturbations ($\kappa=60$), the mean of seed-level median Spearman correlations is $0.988$ (95\% CI $[0.987,0.989]$) and mean top-10\% overlap is $0.906$ ($[0.903,0.910]$). Broader perturbations ($\kappa=12$) reduce these values to $0.943$ ($[0.939,0.947]$) and $0.795$ ($[0.787,0.802]$). Thus, global order remains comparatively stable even while membership near a funding cutoff changes materially.

The PVS exponent shows the same governance effect. Comparing $\alpha=1$ with $\alpha=2$ gives mean Spearman correlation $0.953$ ($[0.953,0.954]$) but mean top-10\% overlap only $0.774$ ($[0.768,0.780]$). The exponent is therefore not a cosmetic parameter. It changes which work crosses the decision boundary.

\begin{figure}[t]
\centering
\includegraphics[width=0.82\columnwidth]{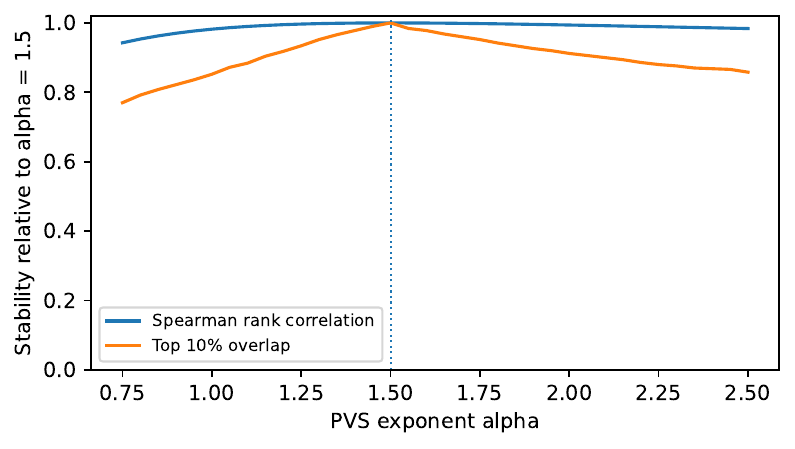}
\caption{Original single-seed ranking sensitivity to the PVS exponent. The 30-seed regeneration reports the same qualitative pattern in the text.}
\end{figure}

\subsection{RQ2: reliability weighting has a correlation-dependent validity region}
At $\rho=0$, the generalized generator closely reproduces the published independent-error result. Inverse-MSE weighting achieves mean MAE $0.615$ (95\% CI $[0.613,0.616]$), a $42.2\%$ improvement over the best individual estimator. Equal weighting reaches MAE $0.635$. Regression-fitted weighting and the oracle ceiling are both approximately $0.614$.

The conclusion changes as error dependence increases. At $\rho=0.4$, inverse-MSE weighting still improves on the best individual by $15.4\%$, while regression-fitted weighting improves by $16.6\%$. At $\rho=0.6$, those gains contract to $4.4\%$ and $7.8\%$. At $\rho=0.8$, inverse-MSE weighting becomes $4.8\%$ worse than the best individual estimator, whereas regression-fitted weighting remains $1.7\%$ better and nearly coincides with the test-set oracle. The regression solution reacts to shared error by concentrating more weight on the strongest marginal estimators rather than treating their errors as independent.

\begin{table}[t]
\caption{MAE under increasing estimator-error correlation. Values are means across 30 seeds.}
\centering\small
\begin{tabular}{lrrrr}
\toprule
Method & $\rho=0.0$ & $\rho=0.4$ & $\rho=0.6$ & $\rho=0.8$ \\
\midrule
Inverse-MSE & 0.615 & 0.902 & 1.017 & 1.115 \\
Regression-fitted & 0.615 & 0.889 & 0.980 & \textbf{1.046} \\
Oracle-test & \textbf{0.614} & \textbf{0.888} & \textbf{0.980} & \textbf{1.046} \\
Equal mean & 0.635 & 0.931 & 1.050 & 1.152 \\
Best individual & 1.063 & 1.065 & 1.063 & 1.064 \\
\bottomrule
\end{tabular}
\end{table}

\begin{figure}[t]
\centering
\includegraphics[width=0.84\columnwidth]{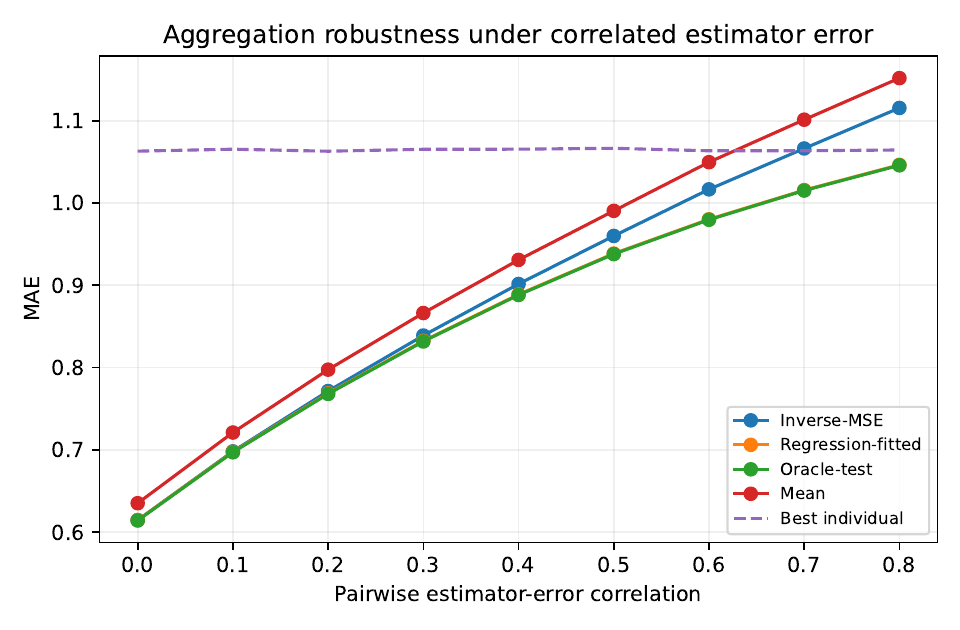}
\caption{Aggregation MAE as estimator errors become positively correlated. Simple inverse-MSE weighting loses its advantage under strong shared error, while calibration-set regression tracks the oracle ceiling more closely.}
\end{figure}

This result narrows the mechanism-level claim. Measured reliability is useful, but marginal reliability alone is insufficient when estimators share failure modes. Under strong dependence, the aggregation rule must account for covariance or fall back toward the strongest estimator.

\subsection{RQ3: adaptive weighting survives seed variation, while CCS weakens under shared noise}
Across 30 seeds, equal weighting has mean Wave-averaged RMSE $0.778$ and four-Wave adaptive reliability weighting $0.742$. The paired relative reduction is $4.64\%$ with 95\% CI $[4.50\%,4.78\%]$. By Wave 20, the mean AI weight rises to $0.403$ ($[0.398,0.408]$), from an initial $0.25$. Stating the rolling window explicitly shows that the adaptation result is not an artifact of an unspecified memory horizon.

\begin{figure}[t]
\centering
\includegraphics[width=0.84\columnwidth]{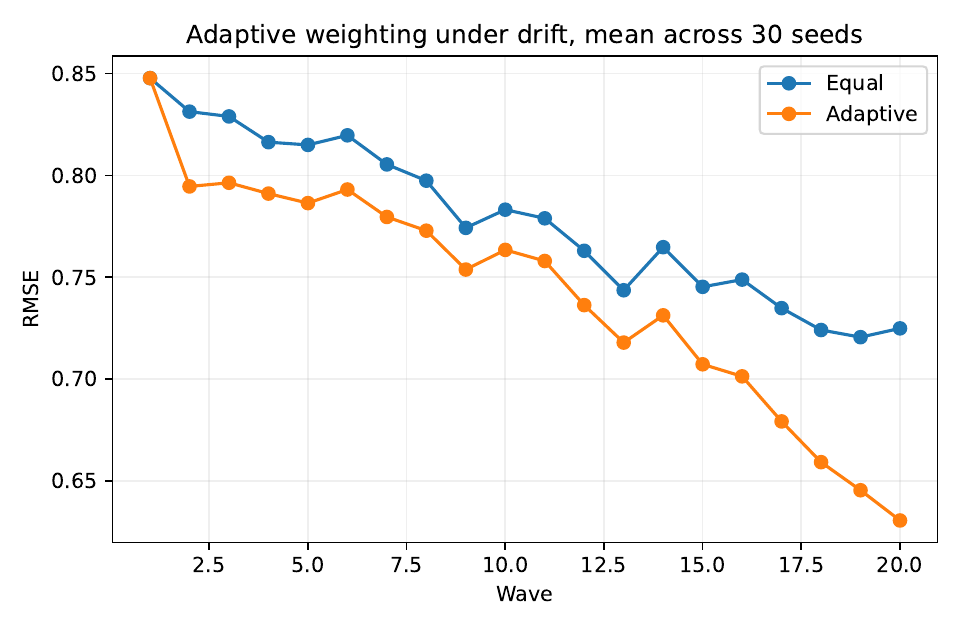}
\caption{Equal and adaptive reliability weighting under drift, averaged over 30 independent seeds. Adaptive weights use the previous four Waves.}
\end{figure}

The CCS diagnostic is strongest when collective information is independent, with mean ROC-AUC $0.908$ ($[0.906,0.910]$) and precision lift $3.39\times$ over prevalence. As shared noise increases, discrimination falls to $0.885$ at $\rho_s=0.4$ and reaches a minimum of $0.879$ ($[0.877,0.881]$) at $\rho_s=0.6$. At $\rho_s=0.8$ AUC is $0.887$. The non-monotonic tail occurs because strongly shared base noise is partly cancelled by differencing, leaving the injected stale-context component relatively more visible. Across the full sweep the diagnostic remains informative, but the experiment confirms that independence assumptions materially affect its apparent strength.

\begin{figure}[t]
\centering
\includegraphics[width=0.84\columnwidth]{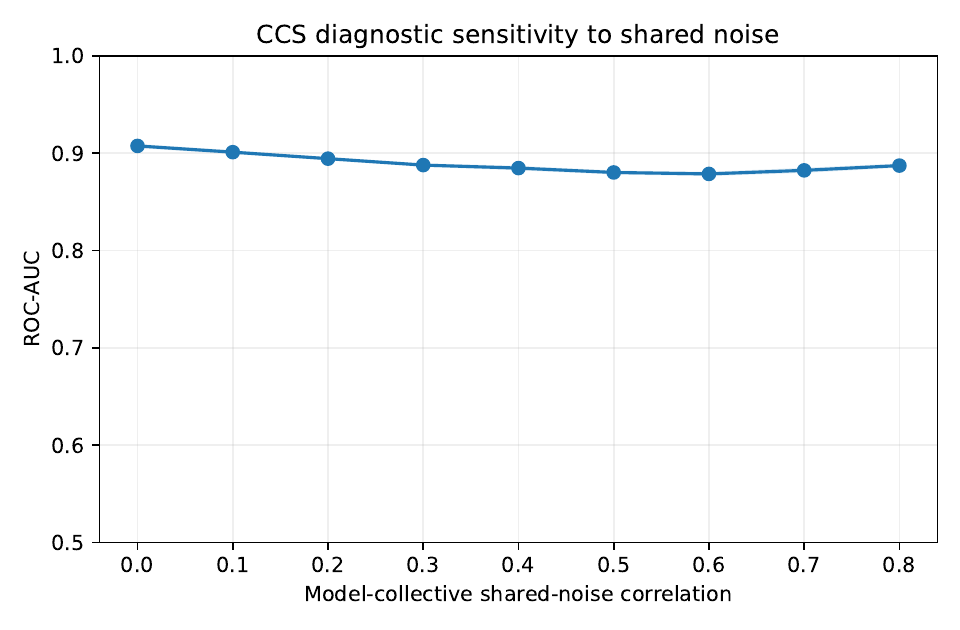}
\caption{CCS diagnostic ROC-AUC as the model and collective channels share increasing base noise. Shading shows 95\% confidence intervals across 30 seeds.}
\end{figure}

\subsection{RQ4: using ODP distance for portfolio selection exposes a boundary}
The original 800-instance benchmark remains unchanged, but its interpretation is now explicit: it repurposes ODP distance as a greedy \emph{release-selection} heuristic. With 12 candidates and exhaustive enumeration, value-to-effort achieves mean objective ratio $0.962$ (bootstrap 95\% CI $[0.958,0.967]$), ODP distance $0.953$ ($[0.948,0.958]$), PVS $\alpha=2$ $0.947$, and BVS greedy $0.933$. The paired value-to-effort advantage over ODP distance is $0.0094$ objective-ratio points ($[0.0038,0.0150]$). This result does not imply that ODP is inferior for dependency-aware sequencing, because the benchmark optimizes subset value under a budget rather than sequence quality.

The new scaling study tests a fresh instance family at 12, 25, 50, and 100 candidates using an exact MILP reference. Value-to-effort remains the strongest transparent heuristic at every size, but its mean ratio declines from $0.966$ at $n=12$ to $0.896$ at $n=100$. ODP distance declines from $0.946$ to $0.849$. The gap between value-to-effort and ODP distance therefore does not disappear at larger $n$ in this generator. More importantly, all heuristic ratios deteriorate as the portfolio grows, confirming that larger dependency-constrained problems create more room for globally suboptimal greedy choices.

\begin{table}[t]
\caption{Objective ratio to the MILP optimum in the scaling study, mean and 95\% CI across 30 seed-level replications.}
\centering\small
\begin{tabular}{lrrrr}
\toprule
Method & $n=12$ & $n=25$ & $n=50$ & $n=100$ \\
\midrule
Value/effort & \textbf{0.966} & \textbf{0.930} & \textbf{0.901} & \textbf{0.896} \\
ODP distance & 0.946 & 0.898 & 0.864 & 0.849 \\
PVS $\alpha=2$ & 0.931 & 0.898 & 0.855 & 0.840 \\
BVS greedy & 0.904 & 0.859 & 0.838 & 0.808 \\
Random & 0.748 & 0.719 & 0.684 & 0.670 \\
\bottomrule
\end{tabular}
\end{table}

\begin{figure}[t]
\centering
\includegraphics[width=0.84\columnwidth]{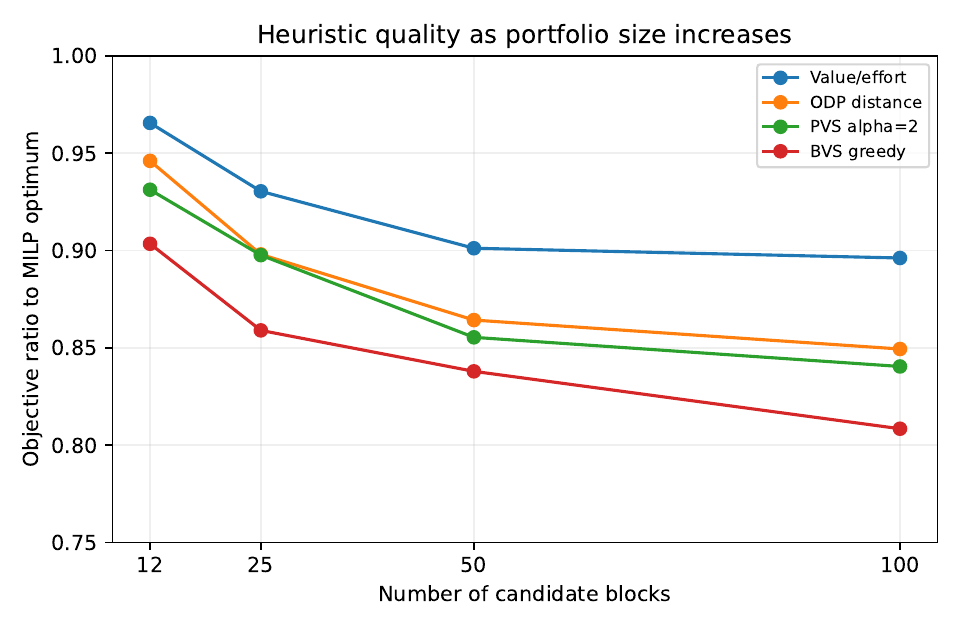}
\caption{Transparent portfolio heuristics against the exact MILP optimum as candidate count increases.}
\end{figure}

For all 30 small-instance cross-checks, the MILP optimum matches independent exhaustive enumeration to floating-point tolerance. This validates the optimization reference before it is used at sizes where $2^n$ subset enumeration is no longer practical.

\subsection{RQ5: ODP is exact without dependencies and near-optimal under the tested prerequisite DAGs}
The native sequencing benchmark reverses the interpretation of the portfolio-selection result. With no prerequisites, ODP's ascending $\mathrm{ERS}/\mathrm{BVS}$ order has efficiency exactly $1.000$ in every replication, as predicted by the ratio-rule derivation. PVS $\alpha=2$ reaches $0.995$, BVS-first $0.978$, ERS-first $0.897$, and random topological ordering $0.823$.

As prerequisites are introduced, the eligible-set ODP policy becomes heuristic but remains close to the exact dynamic-programming optimum. Mean efficiency is $0.986$ at $\lambda=0.5$, $0.982$ at $\lambda=1.0$, and $0.982$ at $\lambda=1.5$. At $\lambda=1.0$, PVS reaches $0.980$, BVS-first $0.971$, ERS-first $0.934$, and random topological ordering $0.896$. The ODP--PVS difference is small under the tested generator, so the result supports near-optimality of the ODP ratio policy for this objective rather than broad superiority over all dependency-aware sequencers.

\begin{table}[t]
\caption{Native sequencing efficiency $J^*/J$ at $n=12$. One is exact optimality.}
\centering\small
\begin{tabular}{lccccc}
\toprule
$\lambda$ & ODP & PVS $\alpha=2$ & BVS & ERS & Random \\
\midrule
0.0 & \textbf{1.000} & 0.995 & 0.978 & 0.897 & 0.823 \\
0.5 & \textbf{0.986} & 0.983 & 0.971 & 0.910 & 0.857 \\
1.0 & \textbf{0.982} & 0.980 & 0.971 & 0.934 & 0.896 \\
1.5 & \textbf{0.982} & 0.982 & 0.978 & 0.953 & 0.931 \\
\bottomrule
\end{tabular}
\end{table}

\begin{figure}[t]
\centering
\includegraphics[width=0.84\columnwidth]{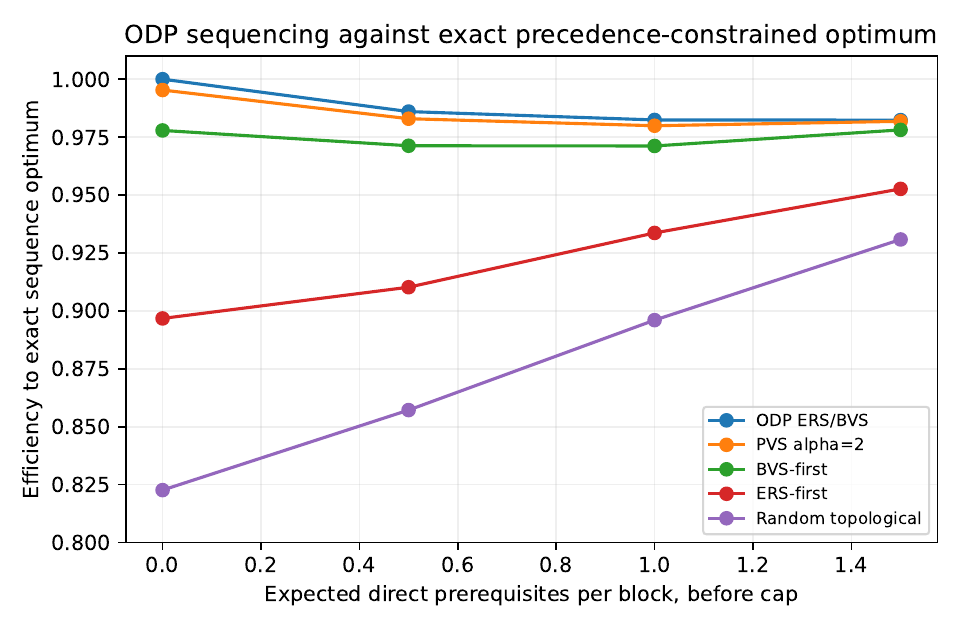}
\caption{Dependency-valid sequence efficiency relative to the exact weighted-completion optimum at $n=12$.}
\end{figure}

The scaling study evaluates early value delivery when all selected blocks must eventually execute. ODP has the highest mean normalized value AUC at each tested size, from $0.508$ at $n=12$ to $0.544$ at $n=100$. PVS is close, from $0.507$ to $0.543$, followed by BVS-first from $0.504$ to $0.539$. At $n=100$, ERS-first reaches $0.518$ and random topological ordering $0.496$. The small ODP--PVS margin again cautions against a universal ranking, but the experiment shows that ODP's native ratio ordering behaves fundamentally differently from using the same distance as a portfolio selector.

\begin{figure}[t]
\centering
\includegraphics[width=0.84\columnwidth]{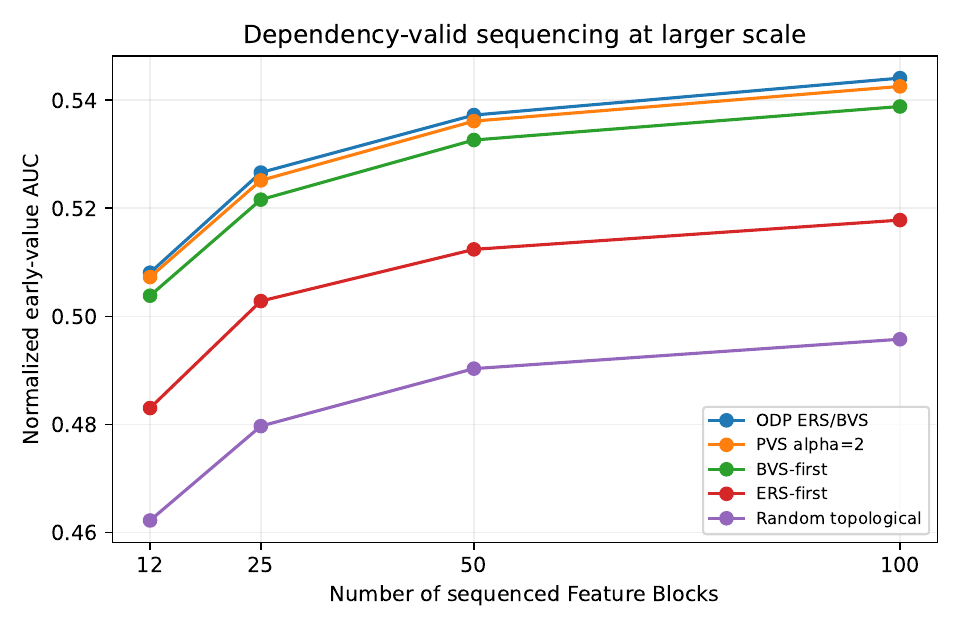}
\caption{Normalized early-value AUC for dependency-valid sequencing at $\lambda=1.0$. Higher is better.}
\end{figure}
\FloatBarrier
\section{Discussion}
\subsection{Prioritization parameters are governance decisions}
The regenerated BVS and PVS studies preserve the original governance result across 30 seeds. High global rank correlation does not imply stable funded-set membership. Parameter ownership, rationale, version history, and replayable rankings are therefore part of the decision evidence, especially near capacity cutoffs.

\subsection{Reliability weighting requires dependence awareness}
The strongest revision concerns estimator aggregation. Under independent errors, inverse-MSE weighting reproduces the large gain reported in the original paper. Under shared error, however, the gain shrinks and ultimately reverses. Marginal accuracy is therefore not enough to determine safe influence. A deployable aggregation layer should measure both estimator reliability and error dependence. The constrained regression baseline is one transparent way to absorb covariance information, and its closeness to the oracle in this experiment suggests that most of the attainable gain can be learned from a finite calibration set. This does not make regression universally optimal. It establishes that correlation is a first-order governance variable when human, analogy, parametric, and AI estimates consume overlapping evidence.

\subsection{Adaptive calibration remains useful, but collective independence is not free}
The drift result survives 30 independent seeds with a narrow confidence interval when the rolling memory is fixed at four Waves. This strengthens the claim that influence can change with observed predictive performance. The CCS extension gives a complementary warning. Model--collective divergence remains informative under substantial shared noise, but its AUC falls as the two channels inherit more common error. Collective judgment should therefore be treated as an additional information channel whose independence must be assessed, not assumed.

\subsection{Selection and sequencing are different optimization problems}
The larger portfolio study preserves the ranking of the transparent \emph{selection} heuristics in the chosen generator, but absolute objective ratios decline as $n$ grows. The practical implication is not that value-to-effort is globally optimal, nor that it replaces ODP. It shows that selecting a subset under budget and prerequisites requires an optimizer matched to that subset objective~\cite{bagnall2001,pitangueira2015}.

The native sequencing study addresses a different question. Once the selected set is fixed, ODP's $\mathrm{ERS}/\mathrm{BVS}$ distance has a direct scheduling interpretation: without prerequisites it is exactly the ratio rule that minimizes the declared BVS-weighted ERS completion burden~\cite{smith1956}. Under prerequisites, the eligible-set version loses its global guarantee but remains within roughly two percent of the exact optimum on average in the tested $n=12$ DAGs. This resolves the apparent contradiction in the earlier portfolio benchmark: ODP distance can be a weaker \emph{selector} while remaining an effective \emph{sequencer}. The two results are complementary, not conflicting.
\section{Scope and Validity}
The claims in this paper are mechanism-level claims supported by controlled synthetic experiments with known latent variables and explicit error processes. The revision broadens those processes specifically where the first open review identified fragile assumptions.

\textbf{Construct scope.} BVS, ERS, PVS, and CCS are parameterized decision operators, not universal measurement scales for business value, effort, risk, or organizational preference. Their numerical behavior is conditional on the variables supplied to them.

\textbf{Internal validity.} The new 30-seed design quantifies run-to-run variation. The correlation study holds marginal estimator moments fixed while varying an equicorrelated common error, which isolates dependence but does not represent every possible covariance structure. The CCS study introduces one explicit shared-noise family and one stale-context mechanism. Other dependence structures, heavy tails, adversarial estimators, or regime changes can be tested with the same protocol.

\textbf{External validity.} The revised experiments do not claim industrial effect sizes. They establish behavior under declared synthetic error models. Real project histories are still required to estimate field prevalence, organizational boundary conditions, human adaptation, and operational costs.

\textbf{Optimization validity.} The original and scaling portfolio studies maximize declared BVS subject to effort and prerequisite constraints. In those studies, ODP distance is deliberately used outside its native sequencing role as one transparent selection heuristic. The scaling generator fixes one dependency process and a 30\% budget fraction. Other graph densities, budget regimes, risk-adjusted objectives, diversity requirements, or multi-period utility functions define different selection problems and may alter heuristic rankings.

\textbf{Sequencing validity.} The weighted-completion objective $J(\pi)=\sum_i \mathrm{BVS}_i C_i^{\mathrm{ERS}}$ is an experimental operationalization of ``earlier value under effort/risk burden,'' not a new canonical D-POAF invariant. ERS is a governed feasibility score, not literal elapsed time. The exact no-dependency ratio result therefore holds for the declared surrogate objective, while calendar time, parallel execution, critical-path duration, architectural coupling, release gates, or different utility functions can require different sequencing models.

\textbf{Oracle interpretation.} The oracle weighting is computed from the test errors and is intentionally non-deployable. It is included only as an empirical ceiling for the stated error model. The small gap between calibration-set regression and the oracle in this study should not be generalized beyond the tested conditions.
\section{Conclusion}
This revision turns the original single-seed mechanism study into a multi-seed robustness analysis. Three conclusions are stronger and more conditional than before. First, governed scoring parameters can preserve broad rank order while changing which work crosses a portfolio cutoff. Second, evidence-weighted estimator influence is valuable only when dependence is part of the evidence: inverse-MSE weighting gives a $42.2\%$ MAE gain under independent errors, but becomes worse than the best individual estimator at $\rho=0.8$, while covariance-aware regression remains close to the oracle ceiling. Third, adaptive weighting remains beneficial under drift across 30 seeds, and model--collective divergence remains discriminative under shared noise although its AUC weakens.

The optimization results now separate two roles that should not be conflated. For budget-constrained release selection, value-to-effort is the strongest transparent heuristic in the tested portfolio generator and using ODP distance as a selector incurs larger regret. For sequencing a fixed selected set, the picture is different: ascending $\mathrm{ERS}/\mathrm{BVS}$ is exactly optimal for the declared weighted-completion objective without prerequisites and remains near the exact optimum under the tested prerequisite DAGs, while leading the transparent sequencing baselines in mean early-value AUC through $n=100$. The engineering principle is therefore not that one D-POAF operator is universally superior. It is that scoring parameters, estimator influence, information dependence, calibration signals, selection objectives, and sequencing objectives should be explicit enough to stress-test against known references and to revise when their assumptions fail.
\section*{Availability and Reproducibility}
The reproducibility archive accompanying this revision contains the complete Python generator for the robustness extension, every deterministic replication seed, generated raw and summary CSV files, all new figure scripts, the original 800-instance portfolio benchmark, the native ODP exact and scaling sequence benchmarks, and the LaTeX manuscript source. The experiments use Python 3.13.5, NumPy 2.3.5, pandas 2.2.3, SciPy 1.17.0, scikit-learn 1.8.0, Matplotlib 3.10.8, and \texttt{numpy.random.Generator(PCG64)}. The archive is prepared for deposition in a persistent research repository. The D-POAF Canonical Specification and Official Terminology remain available through their Zenodo DOIs~\cite{dpoafcanonical,dpoafterminology}, and the Operating Guide documents the practitioner reference mechanisms~\cite{dpoafguide}.

\section*{Competing Interests}
Azzeddine Ihsine and Sara Ihsine are contributors to the D-POAF framework studied in this manuscript and are affiliated with Inovionix, which develops a commercial software platform implementing D-POAF concepts. The D-POAF framework itself is publicly documented. The evaluation protocol includes both a boundary test in which ODP distance is not the strongest portfolio-selection heuristic and a native sequencing test in which the ratio rule is exact without prerequisites and near-optimal under the tested precedence constraints. The authors declare no other known competing financial or non-financial interests that could have influenced the work reported here.

\section*{Acknowledgements}
The authors thank Evgenii Arsentev for the open PREreview that motivated the correlation, multi-seed, shared-noise, scaling, reproducibility, disclosure, and selection-versus-sequencing clarification in this revision. The authors also thank the D-POAF community and Inovionix collaborators for contributions to the framework and reference implementation.

\end{document}